\documentstyle[11pt,aaspp4,flushrt]{article}

\newcommand{\etal}{{\it et al.\/}}

\begin{document}

\title{Calibration of the CH and CN Variations Among Main Sequence
Stars in M71 and in M13\altaffilmark{1}}

\author{Michael M. Briley\altaffilmark{2} and Judith G.
Cohen\altaffilmark{3}}

\altaffiltext{1}{Based partially on observations obtained at the W.M.
Keck Observatory, which is operated jointly by the California Institute
of Technology and the University of California}

\altaffiltext{2}{Department of Physics, University of Wisconsin
Oshkosh, Oshkosh, Wisconsin}

\altaffiltext{3}{Palomar Observatory, Mail Stop 105-24, California
Institute of Technology}

\begin{abstract}

An analysis of the CN and CH band strengths measured in a large sample
of M71 and M13 main sequence stars by Cohen (1999a,b) is undertaken
using synthetic spectra to quantify the underlying C and N abundances.
In the case of M71 it is found that the observed CN and CH band
strengths are best matched by the {\it{identical}} C/N/O abundances which
fit the bright giants, implying: 1) little if any mixing is taking
place during red giant branch ascent in M71, and 2) a substantial
component of the C and N abundance inhomogeneities is in place before
the main sequence turn-off. The unlikelihood of mixing while on the
main sequence requires an explanation for the abundance variations
which lies outside the present stars (primordial inhomogeneities or
intra-cluster self enrichment). For M13 it is shown that the 3883\AA~
CN bands are too weak to be measured in the spectra for any
reasonable set of expected compositions. A similar situation exists for
CH as well.  However, two of the more luminous program stars do appear
to have C abundances considerably greater than those found among the
bright giants thereby suggesting deep mixing has taken place on the M13
red giant branch.

\end{abstract}


\section{INTRODUCTION}

By virtue of their large populations of coeval stars, the Galactic
globular clusters present us with a unique laboratory for the study of
the evolution of low mass stars.  The combination of their extreme
ages, compositions and dynamics also allows us a glimpse at the early
history of the Milky Way and the processes operating during its
formation. These aspects become even more significant in the context of
the star-to-star light element inhomogeneities found among red giants
in every cluster studied to date. The large differences in the surface
abundances of C, N, O, and often Na, Mg, and Al have defied a
comprehensive explanation in the three decades since their discovery.

Proposed origins of the inhomogeneities typically break down into two
scenarios: 1) As C, N, O, Na, Mg, and Al are related to proton capture
processes at CN and CNO-burning temperatures, material cycled through a
region above the H-burning shell in evolving cluster giants may be
brought to the surface with accompanying changes in composition. While
standard models of low mass stars do not predict this ``deep mixing'',
several theoretical mechanisms have been proposed (e.g., the meridional
mixing of Sweigart \& Mengel 1979, and turbulent diffusion, Charbonel
1994, 1995) with varying degrees of success. Moreover, there is ample
observational evidence that deep mixing does take place during the red
giant branch (RGB) ascent of metal-poor cluster stars (the works here
are far too numerous to list - see the reviews of Kraft 1994 and
Pinsonneault 1997 and references therein). 2) It has also become
apparent that at least some component of these abundance variations
must be in place before some cluster stars reach the giant branch.
Spectroscopic observations of main sequence turn-off stars in 47 Tuc
(Hesser 1978; Hesser \& Bell 1980; Bell, Hesser, \& Cannon 1983;
Briley, Hesser, \& Bell 1991; Briley \etal 1994, 1996; Cannon \etal\
1998) and NGC 6752 (Suntzeff \& Smith 1991) have shown variations in CN
and CH-band and Na line strengths consistent with patterns found among
the evolved giants of these clusters. The assumption that these low
mass cluster stars are incapable of both deep dredge-up and significant
CNO nucleosynthesis while on the main sequence leads to the possibility
that the early cluster material was at least partially inhomogeneous in
these elements or that some form of modification of these elements has
taken place within the cluster. Suggested culprits include mass-loss
from intermediate mass asymptotic giant branch stars, novae, and
supernovae ejecta (see Smith \& Kraft 1996, and Cannon \etal\ 1998 for an
excellent discussion of these possibilities).

Thus the observed light element inhomogeneities imply that there is
some aspect of the structure of the evolving cluster giants which
remains poorly understood (the deep mixing mechanism), that the early
proto-clusters may have been far less homogeneous, that intermediate
mass stars may have played a greater role in setting the composition of
the present day low mass stars than previously thought, etc. Indeed,
the evidence sited in the reviews above have lead many investigators to
suggest that a combination of processes are responsible, i.e., many
clusters contain star-to-star inhomogeneities established early in
their histories which have been further altered by deep mixing during RGB
ascent. This of course greatly exacerbates the problem as a knowledge
of the composition of the more easily observed bright red giants will
not tell the whole story of their chemical history - one must also
understand the makeup of the main sequence stars.

Cohen (1999a,b) measured CH and CN indices for a large sample of main
sequence stars in each of M71 and M13.  In the case of M71, clear
variations of both indices were seen at a fixed luminosity among the
main sequence stars which are significantly larger than the
observational uncertainties.  Cohen found that these indices are each
to first order bimodal and they are anti-correlated.  There are
approximately equal numbers of CN weak/CH strong and CN strong/CH weak
main sequence stars in M71. At least qualitatively, these patterns are
very similar to those observed among the bright M71 red giants. In the
case of M13, which is metal poor compared to M71 by a factor of
$\sim4$, the observed indices are much weaker, and variability from
star to star was not clearly detected in either the CH or the CN
index.

The purpose of the present paper is to determine the range of C and N
abundances which will reproduce the behavior observed in the M71 main
sequence stars and to compare this with the range in abundances found
on the giant branch of M71 by Briley, Smith \& Claver (2001). For M13,
we examine the detectability of star-to-star variations in the observed
CH and CN indices for the sample of main sequence stars of Cohen if the
behavior of C, N and O is similar to that found among its giants and
attempt to constrain the allowed range of C, N, and O abundances.

The approach adopted here is to represent the sample of globular
cluster main sequence stars by a series of model atmospheres taken from
isochrones appropriate for M13 and for M71 across the relevant
luminosity range. We demonstrate the validity of the models by
comparing colors generated using them to the observed color magnitude
diagrams (CMDs). These models are then used to generate synthetic
spectra from which CH and CN indices are measured. The model C, N, and
O abundances are then adjusted in an attempt to reproduce the
behavior of the observed stars.

\section{THE ADOPTED MODELS AND VERIFICATION OF THEIR VALIDITY}

The first step in our analysis is to demonstrate that the adopted model
points are a satisfactory match to the observed stars in the region of
the main sequence turnoff for M13 and for M71. The isochrones used here
are from the O-enhanced grid of Bergbusch \& VandenBerg (1992).

Assuming a metallicity of $[Fe/H]$ = $-$0.80 for M71 (Sneden \etal\
1994), seven $T_{eff}$/log g points were chosen from the 14 Gyr
$[Fe/H]$ = $-$0.78 isochrone spanning a range in luminosity of $4.7 <
M_V < 3.0$. Model atmospheres were calculated for each of these points
using the MARCS program (Gustafsson \etal\ 1975) which were then fed
into the synthetic spectrum generating program SSG (Bell \& Gustafsson
1978; Gustafsson \& Bell 1979; Bell \& Gustafsson 1989; Bell, Paltoglou,
\& Tripicco 1994) using the linelist of Tripicco \& Bell (1995). Each
synthetic spectrum was initially computed from 3,000 to 12,000\AA~ in
0.1\AA~ intervals and convolved with $V, R, I$ filter curves as
described in Gustafsson \& Bell (1979) and Bell \& Gustafsson (1989) to
yield colors appropriate to each model. These colors are listed in
Tables 1 and 2. Figure 1(a) compares the isochrone colors with the $V$,
$I$ photometry (corrected for reddening) of Stetson (2000) of $\approx$
4000 stars with
$V < 20$ in M71. A distance of 3.9 kpc and a reddening of
E($B-V$) = 0.25 mag was assumed from the on line compilation of Harris
(1996).

A similar approach was followed in the case of M13 where a metallicity
of $[Fe/H]$ = $-$1.51 (Kraft \etal\ 1992) was assumed. Model points were
chosen from the 16 Gyr $[Fe/H]$ = $-$1.48 grid. The corresponding
colors are given in Table 2 and plotted in Figure 1(b) with the $V$,
$I$ photometry of Stetson (2000). For M13 a distance of 7.2 kpc
and a reddening of E($B-V$) = 0.02 was used (Harris 1996), however, an
additional 0.11 magnitudes in $(m-M)_V$ (i.e., a 5\% increase in distance)
was required for the best fit to the observed CMD.

Also plotted in both panels of Figure 1 are the locations of the Keck
program stars in the cluster CMDs. The isochrones and models adopted
for M71 and M13 are shown to be a good match, with the spread about the
isochrone consistent with observational errors in the case of M13. For
M71, the spread is broader, but still consistent with a combination of
observational errors, reddening variations at a level of 10\% of the
total reddening (see Cohen \& Sleeper 1995), and possibly a few
non-members.

\section {THE OBSERVED CN AND CH INDICES}

The I(uvCN) index employed by Cohen (1999a,b) measured the flux removed
by the 3883\AA~ CN feature as a fraction relative to a nearby
single-side continuum bandpass.  Unfortunately, the observed spectra
are not flux calibrated and therefore include the signature of the
instrumental response. The affect of this is to cause offsets in
one-sided indices such as I(uvCN) (although we note that for
comparisons between stars of similar luminosities, this is not an issue
as they all share similar offsets).

To facilitate removal of this signature and for comparisons with
synthetic spectra, we have re-measured the uv CN feature using the
S(3839) index (Norris \etal\ 1981).  Because S(3839) is the logarithm of
the ratio between a comparison bandpass and a CN ``feature'' bandpass,
slopes induced by instrumental response result in constant zero-point
shifts in the indices which can be quantified and removed. For the
present observations, the shift in the index was measured using the
continuum fitting routine of IRAF's ``splot'' package. Each spectrum
was fitted with a third order cubic spline and the S(3839) index
computed from each fit. The resulting offset was 0.180$\pm$0.008 and
0.178$\pm$0.007 for the M71 and M13 spectra, respectively. For the
double sided I(CH) (Cohen 1999a), a much smaller shift of
$-$0.009$\pm$0.003 was found in the case of M71 and $-$0.004$\pm$0.006
in M13. As the fits also include the actual stellar continua, the same
procedure was adopted for the synthetic spectra as well (we are
essentially comparing indices of flattened spectra). One
sigma errors in S(3839) were also calculated from Poisson statistics in
the feature and continuum bandpasses.

\section {THE COMPUTATION OF THE CN AND CH INDICES FROM SYNTHETIC SPECTRA}

For each of the model $T_{eff}$/log g points taken from the isochrones,
high resolution synthetic spectra were generated using the SSG  code.
Scaled solar abundances based on the metallicities above were adopted
with differing
C/N/O abundances (discussed below).  These spectra were computed from
3,000 to 6,000\AA~ at a wavelength step size of 0.02\AA~ and a
microturbulent velocity of 1.5 km/s. They were then convolved with a
Gaussian of width 8\AA~ to match the resolution of the observed
spectra. S(3839) and I(CH) indices were measured from the smoothed
spectra and are given in Tables 1 and 2.  As with the observed spectra,
the smoothed synthetic spectra were imported into IRAF and the
continuum fitted with the splot routine. Indices measured from the
resulting fits yielded a zero point shift of 0.025$\pm$0.001 in S(3839)
for both M71 and M13 spectra.

The C/N/O abundances used in the calculation of synthetic spectra for
M71 are based on abundances used by Briley \etal\ (2001) in a DDO
photometric study of M71 bright giants. Using CN sensitive C(4142) and
CH sensitive C(4245) colors it was found that the locus of CN-weak
giants of M71 were best fit by $[C/Fe]$=0.0, $[N/Fe]$=+0.4,
$[O/Fe]$=+0.4 and by $[C/Fe]$=$-$0.3, $[N/Fe]$=+1.4, $[O/Fe]$=+0.2 for
the CN-strong stars. To aid in comparison with the present
main-sequence stars, their results are shown in Figure 2.

For M13, Smith \etal\ (1996) performed an analysis of CN and CH band
strengths in eleven of the cluster's bright giants. Their resulting
abundances centered roughly about $[C/Fe]$=$-$0.85, $[N/Fe]$=+0.7,
$[O/Fe]$=+0.4 and $[C/Fe]$=$-$1.1, $[N/Fe]$=+1.2, $[O/Fe]$=$-$0.5 for
the CN-weak and strong giants respectively and a set of models with
these compositions have been calculated here. However, the low [C/Fe]
abundances of the evolved giants also suggests a deep mixing process
may have altered the initial C/N/O abundances of these stars. As a test
of this possibility, we have also calculated a set of models assuming
arbitrary ``pre-mixing'' abundances of $[C/Fe]$=0.0, $[N/Fe]$=+0.4,
$[O/Fe]$=+0.4 and $[C/Fe]$=$-$0.3, $[N/Fe]$=+0.9, $[O/Fe]$=0.0 (with
the requirement that the total number of C/N/O atoms remain constant).

\section {DISCUSSION}

\subsection {M71}

The observed CN and CH band strengths for 77 M71 main-sequence stars
are plotted against their brightness in $R$ (see Cohen 1999b) in Figure
3. The stars have arbitrarily been divided into CN-weak and strong
groups based on their S(3839) indices. As was pointed out by Cohen
(1999b), the sample falls into a bimodal distribution with a pronounced
CN/CH anticorrelation. This pattern is evident among the more luminous
giants of the cluster as well (see Figure 2). Moreover, the relative
numbers of CN-weak and strong stars are remarkably similar: for the
bright giants, 48 are CN-weak and 30 are strong. Among the
main-sequence stars, 44 are weak and 31 are strong. Three stars marked
as diamonds in Figure 3 have anomalously large CN indicies. 
Based on radial velocities from higher dispersion spectra to be presented
in a later paper, one of these three stars (star
1951333+184107) is not a member of M71 and star
1951311+183525 is probably not a member. 

Also plotted in Figure 3 are the indices measured from the synthetic
spectra described above using the same set of C/N/O abundances which
fit the giants in Figure 2 (the zero point shifts in the indices have
been applied). The agreement between the fits of the CN and CH
sensitive indices in both the giants and main-sequence stars is
remarkably good, suggesting similar C and N abundances in the two
samples and little change in these abundances over the six magnitudes
observed.

Similar results have also been found among the giants and main-sequence
stars of 47 Tucanae, a nearby cluster of comparable metallicity to M71.
The C/N/O abundances used here are very close to those used by Briley
(2001) to fit the CN and CH band strengths of 283 giants, the only
difference being an $[O/Fe]$ of +0.45 assumed for both CN-weak and
CN-strong giants in 47 Tuc. Among the main-sequence stars,
spectroscopic abundance studies have also yielded C and N abundances
similar to those found among the 47 Tuc giants (see Briley \etal\ 1994
and Cannon \etal\ 1998).

Thus both M71 and 47 Tuc, two relatively metal-rich clusters, appear to
show little evidence of change in C/N/O abundances from the upper
main-sequence through the bright giants. This is in striking contrast
to the significant (factor of ten) C depletions with luminosity seen
among the more metal-poor clusters such as M92, M15, and NGC 6397 (see
references in the reviews mentioned above). While at first appearing
contradictory, the difference may well be the result of the higher
H-burning shell temperatures and smaller mean molecular weight
gradients found in more metal-poor stars (as discussed by Sweigart \&
Mengel 1979, Charbonnel 1995, Cavallo \etal\ 1996, 1998, and others)
which should lead to increasingly significant C/N/O abundance changes
due to dredge-up with decreasing metallicity. The present results then do
not rule out the operation of deep dredge-up in evolving cluster
giants, particularly in metal-poor stars.

However, it is clear the process responsible for the C and N
inhomogeneities in M71 and 47 Tuc cannot be deep mixing during 
the ascent of the RGB: the bulk of the inhomogeneities are in place at the main
sequence turn-off. Moreover, the shallow convective zones and low rates
of CN-cycle processing in low mass cluster main sequence stars make
mixing on the main sequence an unlikely scenario. One is then left with
the possibilities of a primordial inhomogeneity or some pollution
mechanism operating early in the cluster histories. In this regard, M71
provides an interesting data point as it is almost four magnitudes
fainter in luminosity than 47 Tuc (Harris 1996) and would be required
to be considerably more massive in the past (not an unlikely scenario
given M71's low galactic latitude) if the observed inhomogeneities are
indeed due to retained gas from intermediate mass AGB stars.

\subsection {M13}

The situation in M13 is considerably different from M71. With its lower
metallicity, both the hotter main-sequence turn-off and lower numbers
of both C and N atoms conspire to greatly reduce CN formation. This
can be seen in Figure 4 where the S(3839) indices of the program stars
with V photometry of Stetson (2000) are plotted against the indices from
the synthetic spectra. Despite significant differences in the C and N
abundances used in the models, the 3883\AA~ CN band is in all cases
too weak to be detected
and cannot be used to constrain N abundance variations among the
M13 main-sequence stars. There does appear to be some CH formation,
however the band is too weak to be useful at the signal level of our
spectra at depletions of $[C/Fe] \approx -0.3$ or greater at the main
sequence turn-off.

We do note the two more luminous M13 stars (at $R$ = 17.9) which do
apparently exhibit a $\approx$ 0.3 dex difference in $[C/Fe]$. Their C
abundances are also substantially greater than C abundances found among
eleven bright M13 giants by Smith \etal\ (1996) which averaged near
$[C/Fe] = -1.$ This result, although only based on two stars, is
consistent with an episode of deep mixing taking place on the M13 RGB
as has also been suggested by the high resolution Na study of
Pilachowski \etal\ (1996).

\acknowledgements

We are very grateful to Peter Stetson for providing access to his
photometric database. We also wish to express
our thanks to Roger Bell whose SSG code was instrumental in this
project. Partial support was provided by a Theodore Dunham, Jr. grant
for Research in Astronomy and the National Science Foundation under
grant AST-9624680 to MMB and grant AST-9819614 to JGC.

\section {APPENDIX: COORDINATES FOR THE M13 AND M71 MAIN SEQUENCE STARS}

During the course of this work, the coordinates published in Cohen
(1999a,b) for the main sequence stars were checked against a set of
coordinates tied to the USNO APM survey (Monet \etal\ 1999).  For M71,
the published coordinates are within 1.5 arcsec of the astrometric
coordinates.  For M13, the published coordinates for objects in Field 2
are of comparable accuracy.  However, the coordinates published in
Cohen (1999a) for main sequence stars in field 1 of M13 are 0.3 sec of
time West and 7 arc sec South of the astrometric coordinates.

\clearpage

\begin{deluxetable}{ccccccccc}
\tablecaption{Model Points and Resulting Colors for M71
\label{tbl-1}}
\tablewidth{0pt}
\tablehead{
\colhead{$T_{eff}$} &
\colhead{Log g} &
\colhead{$M_V$} &
\colhead{$(V-I)$} &
\colhead{$M_R$} &
\colhead{$I(CH)$\tablenotemark{a}} &
\colhead{S(3839)\tablenotemark{a}} &
\colhead{$I(CH)$\tablenotemark{b}} &
\colhead{S(3839)\tablenotemark{b}}
}
\startdata
6010	& 4.40	& 4.698	& 0.632	& 4.387	& 0.150	& 0.023	& 0.104	& 0.189 \nl
6062	& 4.34	& 4.495	& 0.619	& 4.191	& 0.137	& 0.015	& 0.095	& 0.160 \nl
6084	& 4.28	& 4.295	& 0.613	& 3.994	& 0.131	& 0.011	& 0.091	& 0.145 \nl
6077	& 4.20	& 4.096	& 0.613	& 3.795	& 0.129	& 0.011	& 0.090	& 0.144 \nl
6024	& 4.11	& 3.898	& 0.623	& 3.592	& 0.137	& 0.018	& 0.095	& 0.166 \nl
5813	& 3.96	& 3.700	& 0.672	& 3.369	& 0.180	& 0.062	& 0.124	& 0.276 \nl
5235 & 3.66	& 3.490	& 0.825	& 3.078	& 0.257	& 0.253	& 0.192	& 0.491 \nl
\enddata
\tablenotetext{a}{$[C/Fe]$=0.0, $[N/Fe]$=+0.4, $[O/Fe]$=+0.4}
\tablenotetext{b}{$[C/Fe]$=$-$0.3, $[N/Fe]$=+1.4, $[O/Fe]$=+0.2}
\end{deluxetable}

\begin{deluxetable}{ccccccccccccc}
\tablecaption{Model Points and Resulting Colors for M13
\label{tbl-2}}
\tablewidth{0pt}
\tablehead{
\colhead{$T_{eff}$} &
\colhead{Log g} &
\colhead{$M_V$} &
\colhead{$(V-I)$} &
\colhead{$M_R$} &
\colhead{$I(CH)$\tablenotemark{a}} &
\colhead{S(3839)\tablenotemark{a}} &
\colhead{$I(CH)$\tablenotemark{b}} &
\colhead{S(3839)\tablenotemark{b}} &
\colhead{$I(CH)$\tablenotemark{c}} &
\colhead{S(3839)\tablenotemark{c}} &
\colhead{$I(CH)$\tablenotemark{d}} &
\colhead{S(3839)\tablenotemark{d}}
}
\startdata
6189 &	4.42 &	4.703 &	0.590 &	4.420 &	0.042 &	-0.088 &	0.035 
&	-0.086 &	0.029 &	-0.088 &	0.028 &	-0.087 \nl
6247 &	4.36 &	4.493 &	0.576 &	4.217 &	0.041 &	-0.093 &	0.035 
&	-0.092 &	0.031 &	-0.093 &	0.030 &	-0.093 \nl
6286 &	4.30 &	4.299 &	0.567 &	4.027 &	0.041 &	-0.097 &	0.036 
&	-0.096 &	0.032 &	-0.097 &	0.031 &	-0.097 \nl
6294 &	4.22 &	4.081 &	0.563 &	3.811 &	0.041 &	-0.099 &	0.036 
&	-0.098 &	0.033 &	-0.099 &	0.032 &	-0.099 \nl
6256 &	4.14 &	3.899 &	0.569 &	3.626 &	0.041 &	-0.098 &	0.036 
&	-0.097 &	0.032 &	-0.098 &	0.032 &	-0.098 \nl
6132 &	4.02 &	3.684 &	0.594 &	3.398 &	0.043 &	-0.092 &	0.036 
&	-0.090 &	0.030 &	-0.092 &	0.029 &	-0.091 \nl
5874 &	3.87 &	3.510 &	0.653 &	3.195 &	0.056 &	-0.075 &	0.041 
&	-0.069 &	0.028 &	-0.076 &	0.026 &	-0.074 \nl
5512 &	3.66 &	3.300 &	0.742 &	2.938 &	0.105 &	-0.035 &	0.074 
&	-0.007 &	0.038 &	-0.051 &	0.031 &	-0.040 \nl
5383 &	3.53 &	3.096 &	0.777 &	2.717 &	0.126 &	-0.009 &	0.091 
&	 0.037 &	0.045 &	-0.040 &	0.036 &	-0.020 \nl
\enddata
\tablenotetext{a}{$[C/Fe]$=0.0, $[N/Fe]$=+0.4, $[O/Fe]$=+0.4}
\tablenotetext{b}{$[C/Fe]$=$-$0.3, $[N/Fe]$=+0.9, $[O/Fe]$=+0.0}
\tablenotetext{c}{$[C/Fe]$=$-$0.85, $[N/Fe]$=+0.7, $[O/Fe]$=+0.4}
\tablenotetext{d}{$[C/Fe]$=$-$1.1, $[N/Fe]$=+1.2, $[O/Fe]$=-0.5}
\end{deluxetable}

\clearpage

\begin{figure}
\epsscale{1.2}
\plotone{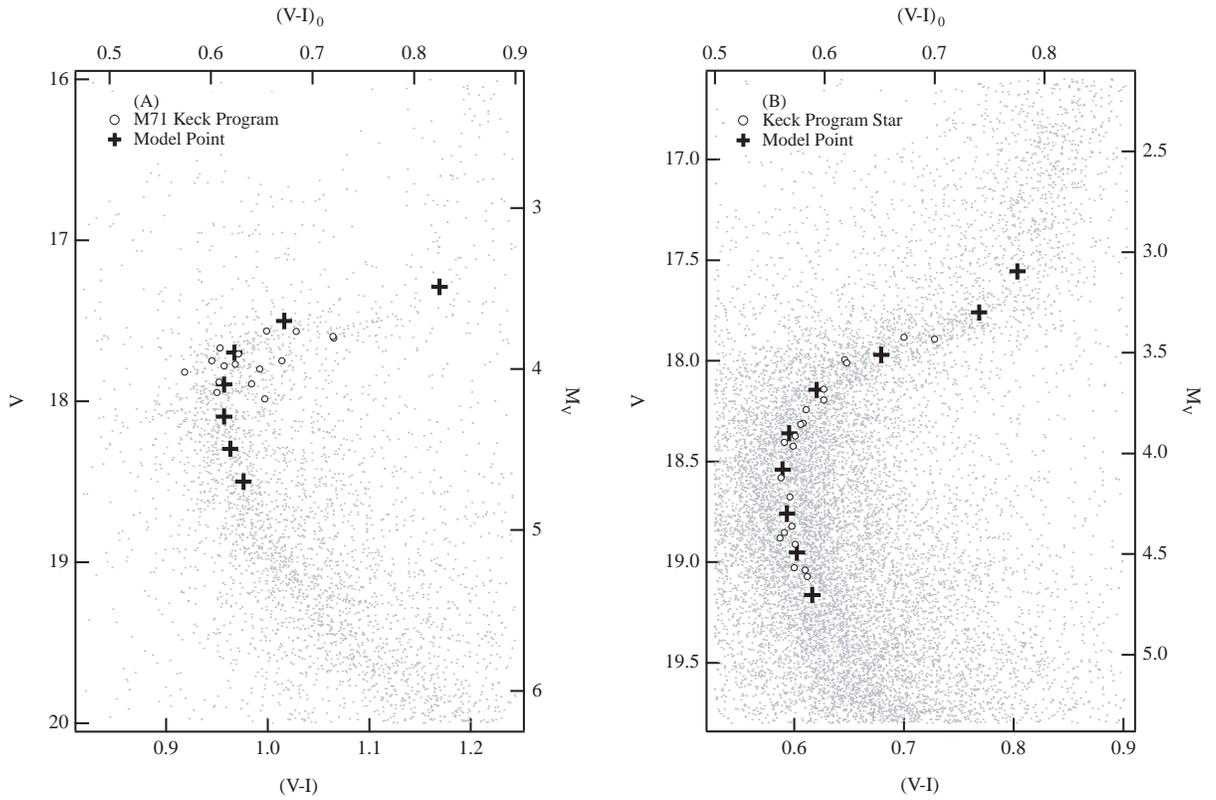}
\caption[Briley.fig1.eps]{The observed CMD's for M71 (a) and M13 (b) from
the photometry of Stetson (2000). Program stars included in his
photometric database are indicated by open circles. Also plotted are the colors
corresponding to the model points listed in Tables 1 and 2.
\label{fig1}}
\end{figure}

\begin{figure}
\epsscale{0.7}
\plotone{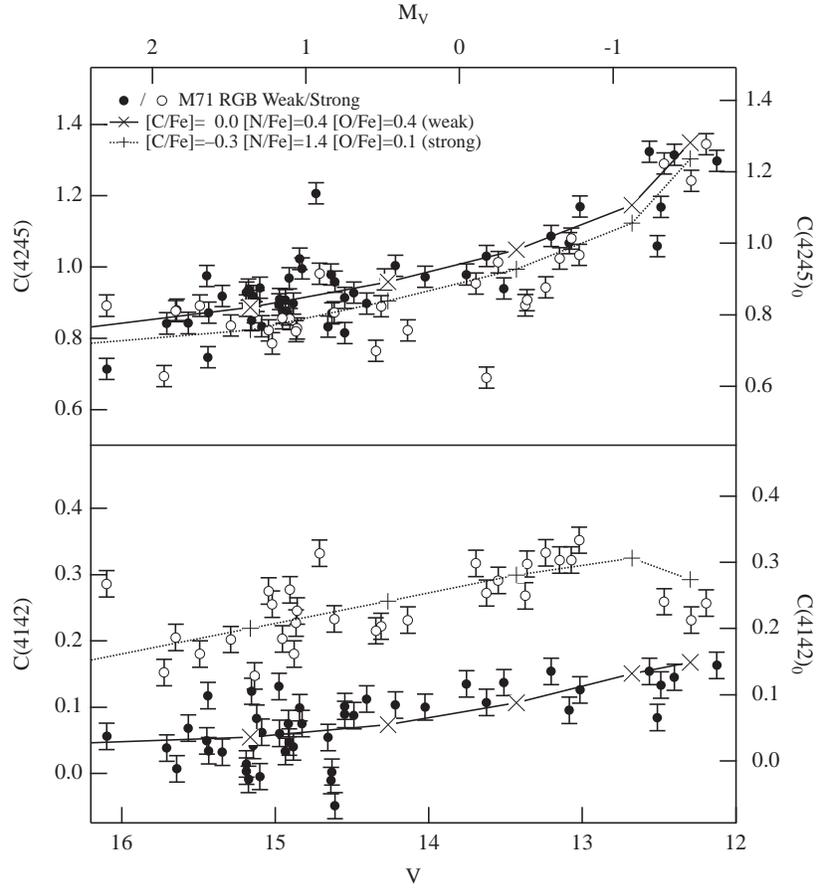}
\caption[Briley.fig2.eps]{The DDO colors for M71 red giants from Briley et
al. 1999. The C(4245) color is sensitive to G-band absorption (CH) and
C(4142) includes the 4215\AA~ CN band. The stars are grouped
according to CN band strength: filled symbols for CN-weak and
open symbols for CN-strong. Synthetic DDO colors for the
C/N abundances chosen for M71 are also plotted.
\label{fig2}}
\end{figure}

\begin{figure}
\epsscale{0.7}
\plotone{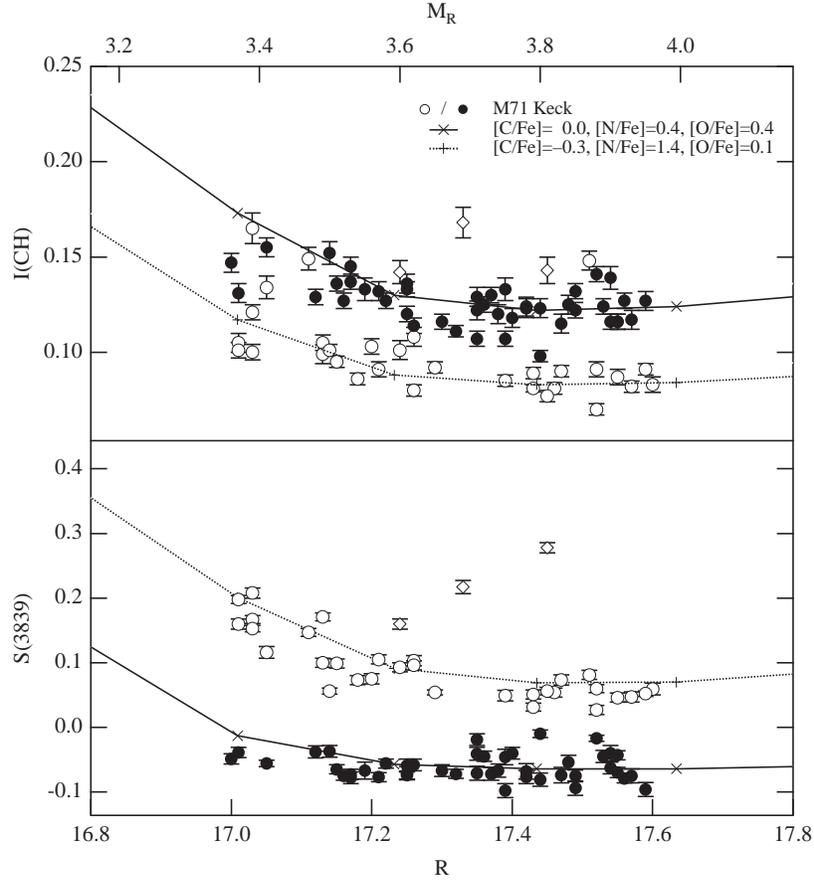}
\caption[Briley.fig3.eps]{The CH and CN indices from M71 main sequence stars
measured from the spectra of Cohen 1999b are plotted against R
magnitude as measured by Cohen. The shifts discussed in the text have
been applied. As discovered by Cohen, the stars appear to naturally
fall into a bimodal distribution with CN and CH largely anti-correlated
(closed symbols are used for CN-weak stars, open for CN-strong). The
three triangles are stars with anomalous CN indices, at least one of
which is a non-member (see text). Also plotted are the indices measured
from the models listed in Table 1 for the two C/N/O compositions used
in Figure 2. Clearly the same composition which fits the evolved M71
stars also fits the present main sequence sample - no modification of
C/N/O during RGB ascent appears necessary.
\label{fig3}}
\end{figure}

\begin{figure}
\epsscale{0.7}
\plotone{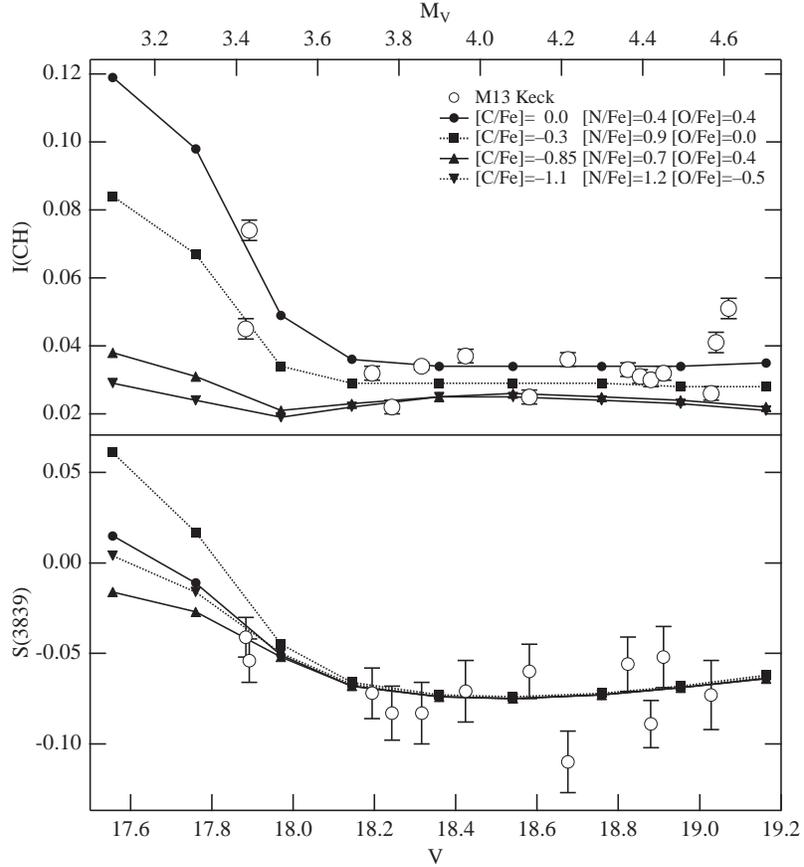}
\caption[Briley.fig4.eps]{The CH and CN indices measured from spectra of M13
main sequence stars by Cohen (1999a) are plotted against V magnitude
from Stetson (2000). The shifts discussed in the text have been
applied. The indices from the model spectra with the C/N/O abundances
discussed in the text are also shown. CN formation is essentially
absent over the entire luminosity range. A similar situation for CH
also exists, although the two most luminous stars appear to have C
abundances considerably higher than those found among the bright giants
by Smith et al. (1996), implying some deep mixing has occurred in M13.
\label{fig4}}
\end{figure}

\end{document}